\documentstyle[prl,twocolumn,aps]{revtex}

\newcommand{\extraspace}{\addtolength{\abovedisplayskip}{2mm} 
                        \addtolength{\belowdisplayskip}{2mm} 
                        \addtolength{\abovedisplayshortskip}{2mm} 
                        \addtolength{\belowdisplayshortskip}{2mm}} 
\newcommand{\be}{\begin{equation}\extraspace} 
\newcommand{\ee}{\end{equation}} 
\newcommand{\bea}{\begin{eqnarray}\extraspace} 
\newcommand{\eea}{\end{eqnarray}} 
 
\newcommand{\nonu}{\nonumber \\[2mm]}

\newcommand{\str}{\rule[-2mm]{0mm}{7mm}}

\newcommand{\STR}{\rule[-7mm]{0mm}{17mm}}

\newcommand{\half}{{\textstyle \frac{1}{2}}} 
\newcommand{\quart}{\frac{1}{4}}

\newcommand{\eps}{\epsilon}

\newcommand{\vac}{| 0 \rangle}

\newcommand{\del}{\partial}

\newcommand{\emp}{\rule[0mm]{0mm}{5mm}}  
\newcommand{\swa}{\displaystyle{\!\!\swarrow\!}}
\newcommand{\nwa}{\displaystyle{\!\!\nwarrow\!}}

\begin{document} 
 
\title{Exclusion statistics for non-abelian quantum Hall states}
\author{Kareljan Schoutens} 
\address{ 
     Institute for Theoretical Physics,
     Valckenierstraat 65, 1018 XE Amsterdam, 
     THE NETHERLANDS} 
\date{March 12, 1998, revised June 8, 1998}
\maketitle
\begin{abstract} 
We determine the exclusion statistics properties of the 
fundamental edge quasi-particles over a specific $\nu=\half$ 
non-abelian quantum Hall state known as the $q=2$ pfaffian 
state. The fundamental excitations are the edge electrons of 
charge $-e$ and the edge quasi-holes of charge $+{e \over 4}$. 
We explicitly determine thermodynamic distribution functions
and establish a duality which generalizes the duality for 
fractional exclusion statistics in the sense of Haldane.

\end{abstract}

\vskip 2mm

\noindent{\small Report number: ITFA-98-04}

\vskip 4mm

\narrowtext

Soon after the discovery of the fractional quantum Hall 
effect, Laughlin proposed the many body wave functions that 
have ever since served as the theoretical basis for the
understanding of the phenomenon. It was quickly established 
that the fundamental quasi-particles over the Laughlin states 
are unusual in a number of ways: in particular, 
they have fractional charge and fractional braiding statistics.

The prediction of fractionally charged excitations has
recently been confirmed in a number of independent experiments 
based on shot noise \cite{noise} and on the physics of the Coulomb 
blockade \cite{Coul}.
The fractional braiding statistics of Laughlin quasi-particles 
are harder to detect in a direct way. Braiding
particles in a two-dimensional system is a theoretical notion
that does not have a direct experimental counterpart.

To explore experimental consequences of fractional {\em braiding}\ 
statistics, one may as a first step try to translate this notion 
into a related notion of fractional {\em exclusion statistics}, i.e., 
into some generalization of the Pauli exclusion principle 
\cite{Ha,FES}. The exclusion statistics lead to a generalization of 
the Fermi-Dirac distribution, and from this one derives predictions 
for observable quantities.

In recent work \cite{Sc,ES}, we have carried out part of the above 
program for the quasi-particles over the $\nu={1 \over m}$ Laughlin 
state. In particular, we have linked the braiding statistics of 
these excitations to the exclusion statistics of the edge 
quasi-particles in the same system. The connection is 
established 
with the help of a Conformal Field Theory (CFT) whose conformal blocks
reproduce the wave functions for the quasi-particle excitations
and whose finite size spectrum coincides with the spectrum of edge
excitations. 
The braiding statistics of excitations over the Laughlin state 
are abelian with statistics angle $\theta = {\pi\over m}$
for the charge $\pm {e \over m}$ excitations and $\theta=m\pi$ 
for the excitations of charge $\pm e$ \cite{MR}.
In \cite{Sc,ES} we established that the exclusion statistics
of edge electrons (of charge $-e$) and edge quasi-holes 
(of charge $+{e \over m}$) are given by fractional 
exclusion statistics in the sense of Haldane \cite{Ha}, with 
statistics parameter $g=m$ and $g={1 \over m}$, respectively, 
and no mutual statistics between the two. This result was 
established by a method based on recursion relations for truncated 
conformal spectra \cite{Sc}.

For a quantum Hall state with fundamental charged edge 
excitations of charge $q$  (measured in units of $e$) and 
distribution function $n(\eps)$ (generalizing the Fermi-Dirac 
distribution), the zero temperature Hall conductance 
is quickly shown to be
\be
\sigma_H = q^2 \, n^{\rm max} \, {e^2 \over h}
\label{sigmaH}
\ee
with $n^{\rm max}$ the maximum (reached for $\eps\to -\infty$)
of $n(\eps)$. For Haldane $g$-ons one has $n_g^{\rm max}={1 \over g}$
and this gives the familiar result $\sigma_H={1 \over m} {e^2 \over h}$, 
both for the edge electrons and the edge quasi-holes over the 
$\nu={1 \over m}$ 
Laughlin state. In this quasi-particle formalism, the fact that 
the Hall conductance does not depend on temperature comes 
about as a consequence of 
the duality between the statistics $g$ and ${1 \over g}$ 
distribution functions \cite{dual}.   

In the present Letter, we extend the above reasoning to the case 
of a quantum Hall state (the so-called $q=2$ pfaffian state) with 
excitations that obey non-abelian braiding statistics. In this 
context, `non-abelian' means that the wave function for a set of 
quasi-particle excitations at given positions has more than one 
component, and that the action of braiding (exchanging) two of those 
excitations is represented by a non-diagonal matrix acting on the 
multi-component wave function. 

The issue addressed in this Letter is a fundamental one: we are 
exploring the consequences for thermodynamics and transport of the 
non-abelian braiding statistics of a set of fundamental excitations 
in a condensed matter system. To our knowledge, a similar problem 
has not been addressed until now \cite{spinS}.
As a characteristic feature of `non-abelian exclusion statistics',
we shall identify a non-trivial pre-factor in the high-energy 
Boltzmann tail of a generalized distribution function. 

After the experimental observation \cite{fivehalves} of a quantum Hall 
effect at the even denominator filling fraction $\nu={5 \over 2}$, 
a number of candidate ground states with filling
fraction $\nu=\half$ have been proposed. One of these is the 
so-called $q=2$ pfaffian state \cite{MR,GWW} with wave function
\be
\Psi_{\rm Pf} = 
{\rm Pf}\left( {1 \over z_i-z_j} \right)
\prod_{i>j} (z_i-z_j)^2 \prod_i e^{-{|z_i|^2 \over 4\ell^2}}
\label{wf}
\ee
with ${\rm Pf}\left( {1 \over z_i-z_j} \right)
= {\cal A}\left({1 \over z_1-z_2}{1 \over z_3-z_4} \ldots \right)$ 
denoting the anti-symmetrized product over pairs of electrons and 
with $\ell=\sqrt{{\hbar c \over eB}}$ the magnetic length.
Following \cite{MR}, one may interpret 
this state as a conformal block of $n$ chiral vertex operators in a 
$c={3 \over 2}$ CFT. In the same manner, the wave functions for electron 
and quasi-hole excitations over the pfaffian state are obtained from 
conformal blocks with the appropriate insertions of electron and 
quasi-hole operators. To write these operators, we represent the CFT 
as a product of a $c=\half$ (Ising) CFT with Majorana fermion $\psi$
and spin operator $\sigma$ and a $c=1$ gaussian theory with scalar 
field $\varphi$. In terms of these fields, 
the electron operator $G(z)$ (of charge $-e$)  
and the quasi-hole operator $\phi(z)$ (of charge +${e \over 4}$)  
take the following form
\bea
G(z) \equiv 
\left( \psi e^{-i \sqrt{2}\varphi} \right)(z) \ ,
\quad
\phi(z) \equiv 
\left( \sigma \, e^{ i {1 \over 2\sqrt{2}} \varphi} \right)(z) \ .
\eea

The electron and quasi-hole operators play a double role in the theory. 
Via their conformal blocks they give bulk wave functions for excited 
states. At the same time, they act as spectrum generating fields in the 
edge CFT. In terms of statistics, the implications of the underlying
CFT are again two-fold. On the one hand, the braiding
statistics in the CFT directly translate into braiding statistics of 
electron and quasi-hole (bulk) excitations. On the other hand, the exclusion 
statistics properties of the electron and quasi-hole operators (see below) 
dictate the thermodynamic (and, to some extent, transport)  properties 
of the low-energy edge theories. For the pfaffian state, the (non-abelian) 
braiding statistics were studied in great detail in \cite{MR,NW2}. The 
study of the exclusion statistics aspects is the subject of this Letter.

The statistics of quasi-particles over the $\nu={1 \over m}$ Laughlin 
states are abelian and can be inferred from the properties of 2-particle 
states. As such, they can be studied by inspecting the short distance 
operator product expansions (OPE) of the electron and quasi-hole fields. 
The OPE's lead to abelian braiding statistics, with angles $m\pi$ and 
${\pi \over m}$ for the electrons and quasi-holes, resp. The 
corresponding exclusion statistics are of the type proposed by Haldane, 
with statistics parameters $g=m$, $g={1 \over m}$, respectively \cite{ES}. 
The OPE's for the electron and quasi-hole edge excitations over the
the pfaffian state have a more complicated structure, and it is no longer
possible to derive the statistics from here. Indeed, the non-abelian 
nature of the braiding statistics of the quasi-hole excitations only
becomes apparent at the 4-particle level.

In \cite{NW2}, it was argued that the wave-function specifying 
a collection of $2n$ quasi-holes (at given locations) over the 
pfaffian state has $2^{n-1}$ independent components. 
In the thermodynamic limit, this degeneracy amounts to a degeneracy of 
$\sqrt{2}$ per quasi-hole. This non-integer degeneracy factor,
which ends up as a pre-factor in the Boltzmann tail of the generalized 
distribution function, is a manifestation of the non-abelian 
nature of the quasi-hole exclusion statistics.

To derive quantitative results for the exclusion statistics of the
edge excitations over the pfaffian state, we employ 
the method of truncated conformal 
spectra \cite{Sc}. The idea of this method is to establish an 
explicit quasi-particle basis of the CFT chiral finite size 
spectrum, and from there derive the statistics. In \cite{ES}, 
this procedure was carried out in great detail for the Laughlin
states and for the charged sector for more general hierarchical 
quantum Hall states (Jain series). 

The complication that is new for the case of non-abelian quantum 
Hall states is the occurrence of so-called fusion channels in the 
specification of a multi-$\phi$ state. These have their origin
in the following fusion rule for the spin field of the Ising CFT
\be
[\sigma] \cdot [\sigma] = [{\bf 1}] + [\psi] \ .
\label{fusion}
\ee
Each time we apply a $\phi$-mode to a state containing an odd
number of $\phi$-quanta, we have to specify a choice for either
the first or the second term on the r.h.s. of (\ref{fusion}).
For a state containing $N$ quasi-holes, the total fusion path 
is encoded in a Bratteli diagram, which is an arrow-path on 
the following grid
\be
\begin{array}{ccccccc} 
 \nwa & \swa & \ldots & \nwa & \swa & \nwa &      \\
 \swa & \nwa & \ldots & \swa & \nwa & \swa & \nwa \\[2mm]
  N   & .\, . & \ldots &  4   &  3   &   2  &  1  
\end{array}
\quad
\begin{array}{c} \psi \\ \sigma \\ {\bf 1} \\ \emp\end{array}
\ee
with the vertical direction indicating the Ising sectors ${\bf 1}$, 
$\sigma$ and $\psi$ and the horizontal index $i=1,\ldots,N$ labeling the
$\phi$-modes. The starting point of an arrow path is the CFT
vacuum $\vac$. The multi-$\phi$ states for a given fusion path are
(with $\phi(z) = \sum_s \phi_{-s} z^{s - 1/8}$)
\be
 \phi_{-{1 \over 8}-\Delta_N -n_N} \ldots 
 \phi_{-{1 \over 8}-\Delta_2-n_2} \phi_{-{1 \over 8}-\Delta_1-n_1} 
 | 0 \rangle 
\ee
with integers $n_N\geq \ldots n_2 \geq n_1\geq 0$ and with
increments $\Delta_i$ determined by the Bratteli diagram
according to

\vspace{3mm}

\begin{tabular}{lcccc}
$\STR \Delta_{i+1} = \Delta_i + \half \quad $ & for \quad &
$ \begin{array}{rl}
  \nwa &       \\
       & \nwa  \\
  i+1 \! &   i   \end{array}$
& and &
$ \begin{array}{rl}
       &  \swa    \\
  \swa &          \\
   i+1 \! &  i \end{array}$ \quad
\\[3mm]
$\STR \Delta_{i+1} = \Delta_i + {1 \over 4} $ & for &
$ \begin{array}{rl}
   \emp  & \emp  \\
   \nwa  & \swa  \\
   i+1 \! &  i \end{array}$
& \hspace{3mm} and & 
$ \begin{array}{rl}
   \swa  & \nwa   \\
   \emp  & \emp   \\
   i+1 \!  &  i \end{array}$
\\[3mm]
$\STR \Delta_{i+1} = \Delta_i  $ & for &
$ \begin{array}{rl}
   \emp  & \emp  \\
   \swa  & \nwa  \\
   i+1 \! &  i \end{array}$
& and &
$ \begin{array}{rl}
   \nwa  & \swa   \\
   \emp  & \emp   \\
   i+1 \!  &  i \end{array}$ 
\end{tabular}

\vspace{3mm}

\noindent 
with $\Delta_1=0$. For example, the lowest energy $4$-quasi-hole states 
for the two possible fusion paths are 
\bea
\begin{array}{cccc}
  \emp &      &      &     \\
  \swa & \nwa & \swa & \nwa 
\end{array}
\qquad
\phi_{-{3 \over 8}}\phi_{-{3 \over 8}}
   \phi_{-{1 \over 8}}\phi_{-{1 \over 8}} \vac \ , \quad E=1 \ , 
\nonumber \\[4mm]
\begin{array}{cccc}
  \emp & \swa & \nwa &     \\
  \swa &      &      & \nwa 
\end{array}
\qquad
\phi_{- {11 \over 8} } \phi_{- {7 \over 8} }
   \phi_{-{5 \over 8}}\phi_{-{1 \over 8}} \vac \ , \quad E=3 \ .
\eea

In the construction of multi-electron states the issue of fusion 
rules does not explicitly arise, but the details of the rules are 
different from what we had for the Laughlin states \cite{ES}. 
On the vacuum $\vac$, we allow the following multi-electron states 
(with $G(z)=\sum_t G_{-t}z^{t-3/2}$)
\be
 G_{-\half-m_{M}} \ldots  G_{-\half-m_2} G_{-\half-m_1} |0\rangle
\ee
with integers $m_i$ satisfying
\be
m_i \geq {\rm max}(m_{i-1}+1,m_{i-2}+4) \ ,
\ee
with $m_0=-2$, $m_{-1}=-3$. For example, the lowest energy multi-$G$
states over $\vac$ are of the form
\be
  \ldots 
  G_{-{13 \over 2}}G_{-{11 \over 2}}G_{-{5 \over 2}}G_{-{3\over 2}}\vac
\ee
with increments alternating between $1$ and $3$.
It is quickly seen that the number of 
states allowed by these rules is larger than for the case with $g=2$, 
where the construction rule is $m_i \geq m_{i-1}+2$.

We are now ready to specify a complete set of independent
multi-particle states. We start by constructing a tower of
states over $\vac$ (which we shall write as $|0,{\bf 1}\rangle$
from now on) by acting independently with $\phi$ and $G$
modes according to the rules that we specified. We then introduce
five additional reference states for which we repeat the procedure.
The lowest $\phi$ and $G$ modes that are allowed on each of the 
reference states are specified in Table 1. 

The central claim of this Letter is that the union of all the quasi-particle 
states in all six sectors precisely forms a basis of the chiral Hilbert space 
of the CFT, and thereby (up to certain projections, see \cite{ES}) of 
physical edge theories. This is a mathematical result which, however, has 
a clear physical interpretation in terms of exclusion statistics (see below). 
The chiral spectrum of the CFT for the pfaffian state is usually understood 
in terms of representations of a $U(1)$ super-affine (Kac-Moody) current 
algebra. The new point of view that we advocate here is very different and 
directly refers to the charge carrying excitations in the system. Using 
Maple, we 
have checked that the character identities that are implied by our claim 
are satisfied to the first 25 non-trivial orders in $q$.

The absence of any interference between 
the construction rules for the $\phi$ and the $G$ quanta means that there 
is no mutual exclusion statistics between the two. This is a first sign of 
a (generalized) particle-hole duality between the two types of 
quasi-particles. We shall make this duality precise in what follows. 

Having specified the quasi-particle basis we can implement
the procedure of \cite{Sc} and derive the thermodynamics.
For the $\phi$-modes, we define truncated and projected
partition sums $X_l(x,q)$ and $Y_l(x,q)$, 
$l={1 \over 8}, {3 \over 8}, \ldots$, as expressions of the form
\be
{\rm trace}_{\leq l} \left( x^{N_{\phi}} \, q^E \right)
\ee
with $N_\phi$ the number of $\phi$-modes, $E$ the dimensionless
energy, $x= e^{\beta\mu_{\rm qh}}$ and $q=e^{-\beta \, {2\pi \over 
L\rho_0}}$ with $\rho_0$ the density of states per unit length.
The subscript $\leq l$ indicates highest occupied mode 
$\phi_{-s}$ is  such that $l-s$ is a non-negative 
integer. $X_l$ ($Y_l$) is further restricted by requiring that
the total charge be an even (odd) multiple of ${e \over 4}$. 
The rules specified above lead to the recursion relations
\bea
&& X_l = X_{l-1} + x \, q^l \, (Y_l + Y_{l-{1 \over 2}}) 
\nonu
&& Y_l = Y_{l-1} + x \, q^l \, X_{l-{1 \over 4}} \ .
\label{recur}
\eea
The initial conditions, which are sector dependent, follow
from the information in Table 1. The solution of (\ref{recur}) 
can be approximated as 
$X_l \sim \prod_{i=1}^l \lambda_+(q^i x)$, with $\lambda_+(z_{\rm qh})$ 
the largest real solution of the equation
\be
(\lambda-1)(\lambda^{1 \over 2} - 1)
 = z_{\rm qh}^2 \, \lambda^{5 \over 4} \ .
\label{lqh}
\ee
This leads to a distribution function
\be
n_{\rm qh}(\eps) = 
  \left[ z_{\rm qh} \del_{z_{\rm qh}} \log(\lambda_+) \right] 
   \left( z_{\rm qh}=e^{ \beta(\mu_{\rm qh}-\eps) } \right)  
\ee
for the quasi-hole excitations.

For the $G$-quanta, the above rules are easily converted 
into a recursion relation for the quantity $\Omega_l(y,q)$,
which is the trace of $y^{N_G} q^E$, with $N_G$ 
the number of $G$ quanta and $y=e^{\beta \mu_{\rm e}}$, over all 
multi-$G$ states with modes $\leq l-\half$. We find
\bea
\Omega_{l} &=& \Omega_{l-1} 
 + y\, q^{l-\half}\, \Omega_{l-2} 
\nonu
&& + y^2 \, q^{2l-2} \, \Omega_{l-4}
   - y^3 \, q^{3l-{13 \over 2}} \, \Omega_{l-6} \ .
\eea
The corresponding distribution function is 
\be
n_{\rm e}(\eps) = 
  \left[ z_{\rm e} \del_{z_{\rm e}} \log(\mu_+) \right] 
   \left( z_{\rm e}=e^{ \beta(\mu_{\rm e}-\eps) } \right) 
\ee
with $\mu_+(z_{\rm e})$ the largest real solution of
\be
(\mu^4-z_{\rm e}^2)(\mu^2-z_{\rm e}) = \mu^5 \ .
\label{le}
\ee

For temperature $T\!=\!0$, the distribution functions
$n_{\rm qh}(\eps)$ and $n_{\rm e}(\eps)$ are step-down 
functions with
maximum $n_{\rm qh}^{\rm max}=8$ and 
$n_{\rm e}^{\rm max}={1 \over 2}$. At finite $T$, the
asymptotic behavior for $\eps\to\infty$ is 
$n_{\rm qh}\sim \sqrt{2} \, e^{-\beta \eps}$ and 
$n_{\rm e}\sim e^{-\beta\eps}$. We stress that the
pre-factor $\sqrt{2}$ is a direct manifestation
of the non-abelian nature of the exclusion statistics
of the quasi-holes.
As an easy consequence of the above formulas,
we establish the duality relation
\be
 2 \, n_{\rm e}(\eps) =
 1 - {1 \over 8} n_{\rm qh}(-{\eps \over 4}) \ , 
\label{dual}
\ee
provided $\mu_{\rm qh}=-\quart \mu_{\rm e}$. This
result generalizes the duality for Haldane's fractional 
exclusion statistics \cite{dual}. The duality allows us 
to interpret the spectrum in different ways.
One picture has quasi-holes with energies ranging from 
$-\infty$ to $\infty$, with the CFT vacuum 
and other reference states represented as filled seas of 
quasi-holes. An alternative is a picture based on edge 
electrons and then there is the ``mixed'' picture that we 
started from, 
with $\phi$ and $G$ quanta of positive energies only.

In the mixed picture, the total central charge
$c={3 \over 2}$ of the edge CFT is established as the 
sum of a contribution $c^+_{\rm qh}$ from the quasi-holes 
and a piece $c^+_{\rm e}$ from the electrons. 
Using Maple, we evaluated
\be
c^+_{\rm qh} = 1.059 \ldots \ , \qquad
c^+_{\rm e} = 0.440 \ldots
\ee
To compute the Hall conductance, we insert 
$q_{\rm qh}={e \over 4}$ and $n_{\rm qh}^{\rm max} = 8$ 
or $q_{\rm e}=-e$ and $n_{\rm e}^{\rm max}= \half$ into
(\ref{sigmaH}), giving $\sigma_H = \half {e^2 \over h}$
as expected. The fact that this result is independent
of temperature is guaranteed by the duality relation
(\ref{dual}). We refer to \cite{ES} for the details
of this way of computing and for further applications
of the quasi-particle formalism.

In conclusion, we have established an appropriate
generalization of the Pauli principle for a situation
where electrons and quasi-holes obey non-abelian
braiding statistics. We have also demonstrated a 
particle-hole duality in this non-abelian theory.
In a future publication, we shall give a more extensive
treatment of exclusion statistics in CFT's with non-abelian 
braiding and give more examples. These will include the
so-called parafermionic quantum Hall states, higher level 
Wess-Zumino-Witten theories and a spinon representation 
for the $SO(5)_1$ CFT, which is relevant for $SO(5)$
invariant ladders models for correlated electrons.

The author thanks P.~Bouwknegt, N. Read and \\
R.A.J.~van Elburg 
for discussions. Part of this work was done at the 1997 
ITP Santa Barbara Workshop on `Low-dimensional Field Theory:
from Particle Physics to Condensed Matter'. This research was 
supported in part by the foundation FOM of the Netherlands.

\vskip 4mm 

\noindent\underline{Note added}
The results presented here are easily generalized to
the more general pfaffian states with factor
$(z_i-z_j)^q$ in (\ref{wf}). The eqs. (\ref{lqh}) and
(\ref{le}) become
\bea
&& (\lambda-1)(\lambda^{1 \over 2} - 1)
 = z_{\rm qh}^2 \, \lambda^{3q-1 \over 2q} 
\nonu 
&& (\mu^{2q}-z_{\rm e}^2)(\mu^{q}-z_{\rm e}) 
 =  \mu^{3q-1} \ .
\eea
and the duality relation reads 
\be
 q \, n_{\rm e}(\eps) =
 1 - {1 \over 4q} n_{\rm qh}(-{\eps \over 2q}) \ .
\ee
The author thanks an anonymous referee for insisting on results for 
general $q$.

After submitting this manuscript, we learned that S.B.~Isakov
[private communication, paper in preparation] has determined 
the exclusion statistics for {\it bulk}\ quasi-holes over the 
general $q$ pfaffian state, and found results that are identical 
to those presented here for {\it edge}\ quasi-holes.

\begin{table}
\begin{tabular}{|c|c|c|c|c|}
\str
\hspace{1.5mm} {\sc State} \hspace{1.5mm} & 
\hspace{1.5mm} {\sc Charge} \hspace{1.5mm} & 
\hspace{1.5mm} {\sc Energy} \hspace{1.5mm} &
\hspace{1.5mm} $\Delta_1$ \hspace{1.5mm} &  
\hspace{1.5mm} $m_{0}$, $m_{-1}$ \hspace{1.5mm} \\
\hline
\str
$ | 0, {\bf 1}\rangle$ & $\!\!$ 0  &  0  & 
  $ 0 $ &
  $ -2, \  -3 $\\
\str
$ | 0 , {\bf \psi}\rangle$ & $\!\!$ 0 & $ \half $ & 
  $ {1 \over 2}$ &
  $ -1, \ -4 $ \\
\str
$ | -{e \over 4},\sigma \rangle$ & $\!\!$ $-{e \over 4}$ & ${1 \over 8}$ &
  $ {3 \over 4}$ &
  $ -1 , \ -3 $ \\
\str
$ | -{e \over 2}, {\bf 1}\rangle$ & $\!\!$ $-{e \over 2}$ & $ {1 \over 4} $ &
  $ {3 \over 4}$ & 
  $ -1 , \  -2 $ \\
\str
$ | -{e \over 2}, {\bf \psi}\rangle$ & $\!\!$ $ -{e \over 2}$ & $ {3 \over 4} $ &
  $ {1 \over 4} $ & 
  $ 0, \ -3 $ \\
\str
$ | -{3e \over 4},\sigma \rangle$ & $\!\!$ $-{3e \over 4}$ & ${5 \over 8}$ &
  $ {1 \over 2}$ &
  $ 0 , \ -2 $ \\
\end{tabular}

\vspace{2mm}

\caption{Reference states for the quasi-particle construction 
of the chiral Hilbert space over the $q=2$ pfaffian state. The 
creation operators $\phi_{-s}$ and $G_{-t}$ act on these states 
according to the exclusion rules specified in the text. 
The symbols ${\bf 1}$, $\psi$ and $\sigma$ denote Ising sectors
and the quantities
$\Delta_1$ and $m_0$, $m_{-1}$ determine the lowest $\phi$ and
$G$ modes that are allowed.}
\end{table}

\end{document}